\documentclass[a4paper,12pt]{article}
\usepackage{hyperref}
\usepackage{amsmath,graphicx}

\usepackage[backend=biber, style=numeric-comp,sorting=nty, giveninits=true]{biblatex}
\usepackage{enumerate}

\usepackage{amsthm}
\usepackage{amsfonts}
\usepackage{amssymb}
\usepackage{subeqnarray}
\usepackage{mathrsfs}
\usepackage{latexsym}
\usepackage{dsfont}
\usepackage{mathtools}
\usepackage{pgfplotstable}
\usepackage{pgfplots}

\usepackage[dvipsnames]{xcolor}
\usepackage{mdframed}
\definecolor{light-gray}{gray}{0.95}

\usepackage[romanian,main=english]{babel}
\usepackage{combelow}

\usepackage{graphicx}
\usepackage{psfrag}

\usepackage{subeqnarray}

\usepackage{amsbsy}

\usepackage{MnSymbol} 
\usepackage{arydshln}

\usepackage[utf8]{inputenc}
\usepackage{nomencl}
\usepackage{xcolor} 
\usepackage{lineno,hyperref}

\usepackage[margin=2cm]{geometry}
\allowdisplaybreaks[4]

\newcommand{\beq}{\begin{equation}}
\newcommand{\eeq}{\end{equation}}
\newcommand{\ba}{\begin{eqnarray}}
\newcommand{\ea}{\end{eqnarray}}

\newcommand{\bs}{\boldsymbol}

\newcommand{\mc}{\mathcal}

\title{Objective kinetic theory for FENE dumbbell suspension}

\author{Liviu Iulian Palade$^1$\thanks{ Corresponding author.  E-mail: liviu-iulian.palade@insa-lyon.fr; giacomin@queensu.ca; giacomin@unr.edu}\, and Alan Jeffrey Giacomin$^{2,}$$^{3,}$$^{4}$    
}

\numberwithin{equation}{section}

\numberwithin{equation}{section}

\begin{document}

\maketitle

\begin{flushleft}

$^1$  Universit\'e de Lyon, CNRS, Institut Camille Jordan UMR 5208\\ INSA-Lyon,  P\^ole de Math\'ematiques, B\^at. Leonard de Vinci No. 401, 21 Avenue Jean Capelle, F-69621, Villeurbanne, France.

$^2$ Physics, Engineering Physics and Astronomy Department, Queen’s University,
Kingston, Ontario K7L 3N6, Canada 

$^3$ Mechanical Engineering Department, University of Nevada, Reno, NV 89557-0312, USA.

$^4$ State Key Lab for Turbulence and Complex Systems, Peking University, Beijing 100871, China

\end{flushleft}

\begin{abstract}

The novelty of this work is that it takes a result of macromolecular theory that is not objective, and fixes it.  To do so we use an objective vorticity tensor to obtain a fully frame invariant form of the classical constitutive equation for FENE dumbbell fluids obtained within the conceptual framework of kinetic theory for polymer solutions.  The influence of such an objective formulation is discussed for steady shear flows.

\end{abstract}

\begin{flushleft}

 

\end{flushleft}


\section{Preamble.}\label{intro}

Based on early ideas set forth by Kirkwood \cite{kirk1968} and by  Bird, Curtiss, Armstrong and Hassager, a kinetic theory of paramount importance for the physics and rheology of polymer solutions and melts was developed, the fundamentals of which are presented in \cite{bird1987dynamics}.  It  since  captured many dominant viscoelastic flow patterns, gaining notoriety.  We recall below only the facts salient to this work. 

This paper is concerned with a particular type of elastic dumbbell suspension, namely with the {\bf F}initely {\bf E}xtensible {\bf N}onlinear {\bf E}xtensible (FENE) connector force between the beads at the centers of which the mass is concentrated.  The general theory is presented in Chap. 13 of \cite{bird1987dynamics}.   At the core of the kinetic theory approach lies a so-called configurational probability $\psi(t,{\bf Q}) $ (equation (13.2-13) of \cite{bird1987dynamics}),

\begin{equation}\label{de1}
 \dfrac{\partial \psi}{\partial t}=-\dfrac{\partial}{\partial {\bf Q}}\cdot \left[ \left( \bs{\kappa} \cdot {\bf Q}\psi \right) -\dfrac{2kT}{\zeta}\dfrac{\partial }{\partial {\bf Q}}\psi -\dfrac{1}{\zeta}{\bf F}^{(c)}\psi \right] 
\end{equation}

\noindent where ${\bf Q}$ is the end-to-end connector vector, ${\bf F}^{(c)}$ the FENE connector force, and $\left[ \bs{\kappa}\right]_{ij} =\dfrac{\partial v_j}{\partial x_i}$ the velocity gradient.  Also, $\bs{\kappa}={\bf L}^T\equiv \left(\nabla  {\bf v}\right) ^T$, with $\nabla  {\bf v}=\dfrac{\partial v_i}{\partial x_j} $ to connect in notations of (\cite{hui1976}, \cite{gurtin1981}, \cite{hui1997}, \cite{joe1990}, \cite{hau2002}, \cite{morro2023}).

Its solution, when obtained explicitly  for particular flows, is key to obtaining the extra stress tensor.  Explicit solutions for general flows are not yet known, but the family of such PDEs itself is known to have well-behaved solutions in the variational sense (\cite{pal2009}, \cite{2010dpde}, \cite{pal2011}, \cite{2012nl}, \cite{2012dcdsb}, \cite{pal2014},  \cite{2014zamp}, \cite{2017ejam}, \cite{2017aa}, \cite{pal2022}, \cite{pal2023}).

Now let $\dot{\bs{\gamma}}:=\bs{\kappa}+\bs{\kappa}^T=2{\bf D} $, $\bs{\omega}:=\bs{\kappa}-\bs{\kappa}^T=-2{\bf W} $ denote the rate of deformation and the vorticity tensor (also called \textit{spin}), respectively.  The $\dot{\bs{\gamma}}$ and $\bs{\omega}$ notations are those of \cite{bird1987dynamics} and e.g.  \cite{lar1988}, and partially of \cite{mk1994}, while ${\bf D}$ and ${\bf W}$ are those of e.g.  \cite{hui1976}, \cite{gurtin1981}, \cite{hui1997}, \cite{hau2002}, \cite{morro2023}.

Consider now the steady state.  First,  sans flow i.e. $\bs{\kappa}={\bf 0} $, an equilibrium solution $\psi_{\text{eq}}({\bf Q})$, is calculated.  Then a solution in the form of an asymptotic expansion - akin to those in  \cite{adel1995}, \cite{kev1996} -   is proposed (see equation (13.5-5) in \cite{bird1987dynamics}),

\begin{equation}\label{ae1}
\psi(\dot{\bs{\gamma}},\bs{\omega},{\bf Q})=\psi_{\text{eq}}({\bf Q})\left[1+\phi_1(\dot{\bs{\gamma}},{\bf Q})+\phi_2(\dot{\bs{\gamma}},\bs{\omega}, {\bf Q})+\dots \right]  
\end{equation}

where functions $\phi_i$ are the solutions of associated ODEs chosen so that when $\bs{\omega}={\bf 0}$ a solution for potential flows is recovered.  Specifically,

\begin{enumerate}[(a)]
 \item 
 
 \begin{equation}\label{phi1}
  \phi_1 = \dfrac{\zeta}{8kT}\dot{\bs{\gamma}}{\bf \colon}{\bf Q}{\bf Q}
 \end{equation}

wherein \cite{bird1987dynamics}  `` ${\bf \colon}$ '' denotes a double contraction over repeated indices (by equation (13.5-10) in \cite{bird1987dynamics}).
 
 \item 
 
 \begin{equation}\label{phi2}
 \phi_2=\left(\dfrac{\zeta}{8kT} \right)^2\left[\dfrac{1}{2} \left( \dot{\bs{\gamma}}{\bf \colon}{\bf Q}{\bf Q}\right)^2 -\dfrac{\langle Q^4\rangle_{\text{eq}}}{15}\dot{\bs{\gamma}}\dot{\bs{\gamma}}+A(Q)\left(\dot{\bs{\gamma}}{\bs{\omega}}{\bf \colon}{\bf Q}{\bf Q} \right)     \right], 
 \end{equation}
 
 with
 
 \begin{equation}\label{phi21}
  A(Q)=\text{const}\left[1-\dfrac{1}{2}\left(\dfrac{Q}{Q_0} \right)^2  \right]
 \end{equation}

 where $Q=|{\bf Q}|$ and $Q_0$ denoting the maximum allowed dumbbell stretch (by equations (13.5-11) and (13.5-13) in \cite{bird1987dynamics}).
\end{enumerate}

Finally the extra stress tensor is obtained using the Giesekus' expression  which involves Oldroyd's upper convected derivative ${\bf Q}{\bf Q}_{(1)}\equiv \overset{\nabla}{\overline{ {\bf Q}{\bf Q}}} $. Thus, from equation (13.5-17) in \cite{bird1987dynamics}, we glean that the polymer contribution to the total extra stress Cauchy tensor depends explicitly on, and only on, tensors $\dot{\bs{\gamma}}$ and ${\bs{\omega}}$.  

We next discuss objectivity and introduce an objective expression for the vorticity tensor.


\section{Remarks on objective tensors}\label{rot}

\subsection{Generalities}\label{rotg}

Recall that the velocity gradient is not objective.  Indeed, for any ${\bf \mc{Q}}\in \mathit{Orth} $, ${\bf L}^{^*}={\bf \mc{Q}}{\bf L}{\bf \mc{Q}}^T +\dot{\bf \mc{{Q}}}{\bf \mc{Q}}^T $ (e.g. equation (12.18) of \cite{hui1976}, equation (10.15) of \cite{hui1997}, or equation (4.49) of \cite{hau2002}).  Consequently, while the strain rate tensor 

\begin{equation}
{\bf D}=\dfrac{1}{2}\left({\bf L}+{\bf L}^T \right) 
\end{equation}

is objective, i.e. 

\begin{equation}
{\bf D}^{^*}={\bf \mc{Q}}{\bf D}{\bf \mc{Q}}^T, 
\end{equation}

the vorticity tensor

\begin{equation}
{\bf W}=\dfrac{1}{2}\left({\bf L}-{\bf L}^T \right) 
\end{equation}

is not:

\begin{equation}
{\bf W}^{^*}={\bf \mc{Q}}{\bf L}{\bf \mc{Q}}^T + \dot{\bf \mc{{Q}}}{\bf \mc{Q}}^T.
\end{equation}

Next, equation (13.5-17) of \cite{bird1987dynamics}  introduces the stress tensor based on arguments proper to the kinetic theory in terms of ${\bf D}^{2}$ and ${\bf W}{\bf D}-{\bf D}{\bf W} $ and of higher order.  While ${\bf D}^{2^*}={\bf \mc{Q}}{\bf D}^2{\bf \mc{Q}}^T  $ is objective, the others are not, as 

\begin{equation}\label{wddw}
{\bf W}^{^*}{\bf D}^{^*}-{\bf D}^{^*}{\bf W}^{^*}={\bf \mc{Q}}\left( {\bf W}{\bf D}-{\bf D}{\bf W}\right) {\bf \mc{Q}}^T -{\bf \mc{Q}}{\bf D}{\bf \mc{Q}}^T{\bf \dot{\mc{Q}}}{\bf \mc{Q}}^T. 
\end{equation}

Moreover, 

\begin{enumerate}[(1)]
 \item it appears from above that a product of an objective tensor with a non-objective tensor results in a non-objective quantity
 
 \item if an arbitrary tensor ${\bf M} $ is objective, ${\bf M}^{^*}= {\bf \mc{Q}}{\bf M}{\bf \mc{Q}}^T$, then its upper convective derivative (essentially a Lie derivative on a smooth manifold) is - classically - also objective for it changes as $\overset{\nabla}{{\bf M^{^*}}}={\bf \mc{Q}}\overset{\nabla}{{\bf M}}{\bf \mc{Q}}^T $  
 
 \item however, the upper convected derivative of a non-objective tensor is still a non-objective tensor.  Consider for example the vorticity tensor ${\bf W} $:  
 
 \begin{eqnarray}\label{wuc}
 \overset{\nabla}{{\bf W^{^*}}}=&&2{\bf \mc{Q}}\overset{\nabla}{{\bf W}}{\bf \mc{Q}}^T + 3\dot{{\bf \mc{Q}}}\dot{{\bf \mc{Q}}}^T +\ddot{{\bf \mc{Q}}}{\bf \mc{Q}}^T +{\bf \mc{Q}}\dot{{\bf \mc{Q}}}^T{\bf \mc{Q}}{\bf L}^T{\bf \mc{Q}}^T-{\bf \mc{Q}}\ddot{{\bf \mc{Q}}}^T \nonumber\\
&&-\dot{{\bf \mc{Q}}}{\bf W}{\bf \mc{Q}}^T- \dot{{\bf \mc{Q}}}{\bf L}^T{\bf \mc{Q}}^T-{\bf \mc{Q}}{\bf W}\dot{{\bf \mc{Q}}}^T-\dot{{\bf \mc{Q}}}{\bf \mc{Q}}^T. 
 \end{eqnarray}

 \item in fact, Oldroyd's lower convected and Jaumann's derivatives of a non-objective tensor lead to non-objective tensors.  
 
 \item thus, put differently, terms formed by products between strain rate tensors $\dot{\bs \gamma} $ or alternatively ${\bf D} $ and the classical vorticity tensor lead to non-objective quantities under the usual operations mentioned just above
 
\end{enumerate}

As it is widely assumed that the stress tensor is objective (e.g. page 87 in \cite{hui1976}, page 129 in \cite{hui1997}, page 170 in \cite{hau2002}, or page 179 in \cite{morro2023}), one is consequently led to explore ways to re-interpret equation (13.5-17) in \cite{bird1987dynamics} so that it becomes properly invariant with respect to objective motions.  We thus next search for other procedures leading to objective quantities involving a suitably chosen spin tensor.


\subsection{Objective vorticity tensors}\label{rovt}

The search for objective vorticity tensors is well documented by VanArsdale \cite{2003vars}.  For instance, Wedgewood \cite{wed1999} found such a tensor to be a solution of a differential equation involving Jaumann's derivative.  Next, based on the work of Truesdell and Noll \cite{tru1992}, VanArsdale defined an objective vorticity tensor via the polar decomposition theorem (and used it to introduce a new family of Rivlin-Ericksen fluids).  Specifically,  let the objective velocity gradient ${\bf L}_D $ be (\cite{2003vars})

\begin{equation}\label{obspin}
 {\bf L}_D={\bf R}\dot{{\bf U}}{\bf U}^{-1}{\bf R}^T={\bf L} - {\bf W}_R, \, {\bf W}_R=\dot{{\bf R}}{\bf R}^T
\end{equation}

\noindent where ${\bf R}\in\mathit{Orth}$ and ${\bf U}\in\mathit{Sym} $ are, respectively,  the rotation and the stretch tensors in the polar decomposition.  As ${\bf L}_D $ is objective - see page 114 of \cite{2003vars} -, so is the associated vorticity tensor  $\bs{\omega}_D=\left({\bf L}_D-{\bf L}_D^T \right) $ to parallel $\bs{\omega}=\bs{\kappa}-\bs{ \kappa}^T $.  One also has:

\begin{equation}\label{potfl}
{\bs \omega}_D=\left({\bf L}_D-{\bf L}_D^T \right)={\bf R}\left(\dot{{\bf U}}{\bf U}^{-1} -{\bf U}^{-T}\dot{{\bf U}}^T\right)  {\bf R}^T
\end{equation}

Now, potential flows are irrotational, so for ${\bf R}={\bf 0} $, from equation \eqref{potfl} one gets ${\bs \omega}_D={\bf 0} $.

\noindent Moreover, since both $\dot{\bs{\gamma}} $  and ${\bs \omega}_D $ are objective, terms linear in  $\dot{\bs{\gamma}}{\bs \omega}_D $ and ${\bs \omega}_D\dot{\bs{\gamma}} $ will be equally objective.  We shall capitalize on this result in the next Section \ref{sce}.  

Consider now a simple shear flow at constant shear rate.  Let $\gamma=\gamma(t) $ be the shear strain and $\dot{\gamma}(t)=\text{const}\equiv \dot{\gamma}$.  Then classically 

\begin{align}\label{shf}
 &  v_x(y)=\dot{\gamma}y & \nonumber\\
 &  v_y=v_z=0 &
\end{align}

Based on the results on page 121 of \cite{2003vars} we get

\begin{eqnarray}\label{ld}
{\bf L}_D=
\left(\begin{array}{ccc}
       0 & \dot{\gamma}-\dfrac{2\dot{\gamma}}{4+\gamma^2} & 0 \\
       \dfrac{2\dot{\gamma}}{4+\gamma^2} & 0 & 0 \\
       0 & 0 & 0 
      \end{array} \right)
\end{eqnarray}

\noindent The usual strain rate tensor is  \cite{bird1987v1}

\begin{eqnarray}\label{gd}
\dot{\bs{\gamma}}=
\left(\begin{array}{ccc}
       0 & \dot{\gamma} & 0 \\
      \dot{\gamma}  & 0 & 0 \\
       0 & 0 & 0 
      \end{array} \right)
\end{eqnarray}

We have thus gathered everything necessary to arrive at a fully objective strain tensor within the framework of the kinetic theory for FENE dumbbell suspension.  We do so presently.

\section{Kinetic theory FENE model }\label{sce}

Equation \eqref{de1} is solved in \cite{bird1987dynamics} up to 2nd order and the solution containing the 0th order term $\psi_{\text{eq}} $, 1st order $\psi_{\text{eq}}\phi_1 $, and 2nd order $\psi_{\text{eq}}\phi_2 $ terms gathered in equation (13.5-15) of \cite{bird1987dynamics}.

\subsection{First order approximation}\label{sce1}

The polymer  contribution to the extra stress tensor $\bs{\tau}_p $ up to the 1st order terms in $\psi$ reads (with the notations of \cite{bird1987dynamics}):

\begin{equation}\label{1stb}
\bs{\tau}_p(\dot{\gamma})=-\dfrac{bnkT\lambda_H}{b+5}\dot{\bs{\gamma}}-\dfrac{b^2nkT\lambda^2_H}{(b+5)(b+7)}\left(\bs{\kappa} \dot{\bs{\gamma}}+\dot{\bs{\gamma}}\bs{\kappa}^T  \right) 
\end{equation}

\noindent which does not appear as such in \cite{bird1987dynamics} but based on it the first row in the right hand side of equation (13.5-17) is finally obtained, for

\begin{equation}\label{kgama}
\bs{\kappa} \dot{\bs{\gamma}}+\dot{\bs{\gamma}}\bs{\kappa}^T=-\dfrac{1}{2}\left(\dot{\bs{\gamma}}\bs{\omega}-\bs{\omega}\dot{\bs{\gamma}}-2\dot{\bs{\gamma}}^2 \right)=-\dfrac{1}{2}\left(\bs{\omega}\dot{\bs{\gamma}}-\dot{\bs{\gamma}}\bs{\omega}-2\dot{\bs{\gamma}}^2 \right) 
\end{equation}

\noindent Using \eqref{ld} and \eqref{gd}, we get the \textit{polymer part} predictions of the two 1st order approximations, namely

\begin{equation}\label{ld1}
\eta(\dot{\gamma})= -\dfrac{\tau_{yx}}{\dot{\gamma}}\,\Rightarrow\, \dfrac{\eta(\dot{\gamma})-\eta_s}{nkT\lambda_H}=\dfrac{b}{b+5}
\end{equation}

\begin{equation}\label{ld2}
\Psi_1(\dot{\gamma})= -\dfrac{\tau_{xx}-\tau_{yy}}{\dot{\gamma}^2}\Rightarrow  \dfrac{\Psi_1(\dot{\gamma})}{nkT\lambda_H^2}=\dfrac{2b^2}{(b+5)(b+7)}
\end{equation}

Next, we propose here to replace the non-objective rate $\bs{\kappa} $ by the objective rate ${\bf L}_D $.  We propose the \textit{least invasive  modifications} to the original theory.  Other possibilities, at this stage of development, might well lead to sound results and we leave this exploration for another day. Consequently, we introduce 

\begin{equation}\label{1stw}
\tilde{\bs{\tau}}_p(\dot{\gamma})=-\dfrac{bnkT\lambda_H}{b+5}\dot{\bs{\gamma}}-\dfrac{b^2nkT\lambda^2_H}{(b+5)(b+7)}\left({\bf L}_D \dot{\bs{\gamma}}+\dot{\bs{\gamma}}{\bf L}_D^T  \right) 
\end{equation}

\begin{equation}\label{gdeta}
 \dfrac{\tilde{\eta}^+(\dot{\gamma})-\eta_s}{nkt\lambda_H} =\dfrac{b}{b+5} \equiv \eta(\dot{\gamma})
\end{equation}

for the steady shear viscosity, and for the first normal stress difference following sudden inception of steady shear flow

\begin{equation}\label{gd2}
\tilde{\Psi}_1^+(\gamma, \dot{\gamma})= -\dfrac{\tilde{\tau}_{xx}-\tilde{\tau}_{yy}}{\dot{\gamma}^2}\Rightarrow \dfrac{\tilde{\Psi}_1^+(\gamma,\dot{\gamma})}{nkT\lambda_H^2}=\dfrac{2b^2}{(b+5)(b+7)}\left(1-\dfrac{4}{4+\gamma^2(t)} \right)>0 
\end{equation}

and 

\begin{equation}\label{gd3}
\dfrac{\tilde{\Psi}_1^+(\gamma,\dot{\gamma})}{nkT\lambda_H^2}\Big|_{\gamma\to0}=0,\, \dfrac{\tilde{\Psi}_1^+(\gamma, \dot{\gamma})}{nkT\lambda_H^2} \Big|_{\gamma\to +\infty} =\dfrac{2b^2}{(b+5)(b+7)} 
\end{equation}

\noindent from which we learn that, for large enough strains, the objective stress produces the same result as the non-objective one.  Put differently, at the 1st order approximation, we do not see any influence in steady shear flow upon operating with an objective stress, even though fundamentally this is necessary.

We next investigate the 2nd order. 

\subsection{Second order approximation}\label{sce2}

The solution to the configurational diffusion equation \eqref{de1} 2nd order approximation is based on an analogy with the rate fluids theory (about which one may consult, among many references e.g \cite{hui1976}, \cite{bird1987v1}, \cite{hui1997}, \cite{joe1990}, \cite{tig1998}, 
\cite{tru2020}, \cite{morro2023}).  The connection between kinetic theory and the rate fluids constitutive laws is discussed on page 80 of \cite{bird1987dynamics}, about which we will recall only the salient elements.  

We again substitute the objective $\bs{\omega}_D $ for the non-objective $\bs{\omega}$ while keeping unaltered all the formal expressions in the right hand side of (13.5-17) of \cite{bird1987dynamics}.  This involves, as is the case in \cite{bird1987dynamics}, lengthy algebra.  After this, one can sum the contributions of the 0th order, 1st order and 2nd order terms.  For steady shear flow, for the polymer contribution to the extra stress tensor, we get

\begin{subeqnarray}\label{ce:ob}
\tilde{\bs{\tau}}_p= & - & \dfrac{bnkT\lambda_H}{b+5}+\dfrac{b^2nkt\lambda_H^2}{2(b+5)(b+7)}\left(\dot{\bs{\gamma}}\bs{\omega}_D-\bs{\omega}_D\dot{\bs{\gamma}}-2\dot{\bs{\gamma}}^2 \right) \slabel{ce:ob1}\\ 
& - & \dfrac{b^3(2b+11)nkt\lambda_H^3}{4(2b+7)(b+5)(b+7)(b+9)}\left[ \bs{\omega}_D\left(\bs{\omega}_D\dot{\bs{\gamma}}-\dot{\bs{\gamma}}\bs{\omega}_D \right)-\left(\bs{\omega}_D\dot{\bs{\gamma}}-\dot{\bs{\gamma}}\bs{\omega}_D \right)\bs{\omega}_D  \right] \slabel{ce:ob2}\\
& + & \dfrac{b^3(2b+11)nkt\lambda_H^3}{4(2b+7)(b+5)(b+7)(b+9)} \left(\bs{\omega}_D\dot{\bs{\gamma}}^2-\dot{\bs{\gamma}}^2\bs{\omega}_D \right) \slabel{ce:ob3}\\
& - & \dfrac{b^3(b+3)nkt\lambda_H^3}{2(b+5)^2(b+7)(b+9)} \text{tr}\left(\dot{\bs{\gamma}}^2 \right)\dot{\bs{\gamma}}\slabel{ce:ob4} 
\end{subeqnarray}

We next calculate the viscometric functions associated with the sudden inception of steady simple shear flow: 

\begin{subeqnarray}\label{vs}
\dfrac{\tilde{\eta}^+(\gamma,\dot{\gamma})-\eta_s}{nkT\lambda_H}  & = &  \dfrac{b}{b+5} \bigg\{ 1+ \dfrac{b}{b+7}\left(\lambda_H \dot{\gamma} \right)  - \dfrac{b^2}{(b+7)(b+9)}\bigg[ \dfrac{2b+11}{2b+7}\dfrac{\gamma^4}{\left(\gamma^2 +4 \right)^2} \nonumber\\
& - & \dfrac{b+3}{b+5}\bigg]\left(\lambda_H \dot{\gamma} \right)^2 \bigg\}
 + \dots \slabel{vs:2}
\end{subeqnarray}

with

\begin{subeqnarray}\label{vs3}
& & \dfrac{\tilde{\eta}^+(\gamma,\dot{\gamma})-\eta_s}{nkT\lambda_H}\Big|_{\gamma\to 0}  = \dfrac{b}{b+5} \left[  1+ \dfrac{b}{b+7}\left(\lambda_H \dot{\gamma} \right) + \dfrac{b^2(b+3)}{(b+5)(b+7)(b+9)}\left(\lambda_H \dot{\gamma} \right)^2\right]+ \dots \slabel{vs3:1}\\ 
& & \dfrac{\tilde{\eta}^+(\gamma,\dot{\gamma})-\eta_s}{nkT\lambda_H}\Big|_{\gamma\to +\infty}= \dfrac{b}{b+5} \left[1+{\color{RoyalBlue}\dfrac{b}{b+7}\left(\lambda_H \dot{\gamma} \right)} - \dfrac{2b^2{\color{ForestGreen}(4b-17)}}{(2b+7)(b+5)(b+7)(b+9)}\left(\lambda_H \dot{\gamma} \right)^2\right]+ \dots \slabel{vs3:2}\nonumber\\
\end{subeqnarray}

\noindent with \eqref{vs3:1} and \eqref{vs3:2} we understand to be new.


The term in blue in \eqref{vs3:2} above is due to working with an objective $\bs{\omega}_D $, and does not appear in equation (13.5-24) in \cite{bird1987dynamics}.  As an aside: in the third term on the right, the \textcolor{ForestGreen}{\bf factor} ${\color{ForestGreen}{\bf (4b-17)}}$ corrects the  $(4b+17)$ of equation (13.5-24).  

Next,

\begin{equation}
 \dfrac{\tilde{\Psi}_1^+(\gamma, \dot{\gamma})}{nkt\lambda_H^2}=\dfrac{2b^2}{(b+5)(b+7)}\frac{\gamma^2}{\gamma^2+4}>0
\end{equation}

as it must, and 

\begin{subeqnarray}\label{nsob}
& & \dfrac{\tilde{\Psi}_1^+(\gamma, \dot{\gamma})}{nkt\lambda_H^2}\Big|_{\gamma\to 0}=0,\slabel{nsob:1}\\
& & \dfrac{\tilde{\Psi}_1^+(\gamma, \dot{\gamma})}{nkt\lambda_H^2}\Big|_{\gamma\to +\infty}=\dfrac{2b^2}{(b+5)(b+7)}+\dots \slabel{nsob:2}
\end{subeqnarray}

We find that:

\begin{enumerate}[(i)]
 
 \item the canonical equation (13.5-25) of \cite{bird1987dynamics} contains \eqref{nsob:2} above as a 1st order term, and we see no objectivity influence at this order for the configurational probability density upon $\tilde{\Psi}_1^+ $.  It is however expected that higher order terms will impact the corresponding results.
 
 \item equation (13.5-25) contains an additional $\sim \left(\lambda_H \dot{\gamma} \right)^2 $ term which is likely due to a higher order term $\phi_3$ in the configurational  probability asymptotic expansion \eqref{ae1}, that is neither calculated in this work nor is it mentioned in \cite{bird1987dynamics};  we leave its daunting calculation for another day

 \item for the second normal stress difference following sudden inception of steady shear flow \cite{bird1987dynamics}, we get new equation $\tilde{\Psi}_2^+(\gamma, \dot{\gamma})=-\dfrac{\tilde{\tau}_{yy}-\tilde{\tau}_{zz}}{\dot{\gamma}^2} $:

\begin{equation}\label{ns2ob1}
 \dfrac{\tilde{\Psi}_2^+(\gamma, \dot{\gamma})}{nkT\lambda_H^2}=-\dfrac{b^2}{(b+5)(b+7)}\dfrac{\gamma^2}{\gamma^2+4}<0
\end{equation}

with

\begin{equation}\label{ns2ob2}
 \dfrac{\tilde{\Psi}_2^+(\gamma, \dot{\gamma})}{nkT\lambda_H^2}\Big|_{\gamma\to 0}=0
\end{equation}

\begin{equation}\label{ns2ob3}
 \dfrac{\tilde{\Psi}_2^+(\gamma, \dot{\gamma})}{nkT\lambda_H^2}\Big|_{\gamma\to +\infty}
=-\dfrac{b^2}{(b+5)(b+7)} 
\end{equation}

\end{enumerate}

which we believe to be new.

\section{Final remarks}\label{fr}

An objective vorticity (spin) tensor $\bs{\omega}_D $, previously employed by VanArsdale to generalize a Rivlin-Ericksen fluid model \cite{2003vars}, is used to obtain an objective constitutive equation for the FENE dumbbell suspension of the kinetic theory.  The exemplary consequences are studied for a simple shear flow at constant shear rate.  Shear viscosity and 1st and 2nd order normal stress differences are calculated and compared to those obtained initially in \cite{bird1987dynamics}.  The calculations are carried out up to the 2nd order in the configurational probability density function.

The main finding is illustrated in equation \eqref{vs3:2}, in which a \textcolor{RoyalBlue}{{\bf term}} $\color{RoyalBlue}{  {\bf \sim} \left({\bs \lambda}_{\bf H}  \dot{{\bs \gamma}} \right)}$, absent from the original result of \cite{bird1987dynamics}, here appears and imparts objectivity to the canonical constitutive equation for the FENE dumbbell suspension. 

We leave the intriguing extensions to other flow types, such as large amplitude oscillatory shear flow or simple extension, for another day.  We close by observing that a long-awaited use for Van Arsdale's objective vorticity tensors \cite{2003vars}, is to take a result of macromolecular theory that is not objective, and to fix it.




\printbibliography

\end{document}